# Internal-Illumination Photoacoustic Tomography Enhanced by a Graded-scattering Fiber Diffuser

Mucong Li, Tri Vu, Georgy Sankii, Brenton Winship, Kohldon Boydston, Russell Terry, Pei Zhong, Junjie Yao

*Abstract* — The penetration depth of photoacoustic imaging in biological tissues has been fundamentally limited by the strong optical attenuation when light is delivered externally through the tissue surface. To address this issue, we previously reported internal-illumination photoacoustic imaging using a customized radial-emission optical fiber diffuser, which, however, has complex fabrication, high cost, and non-uniform light emission. To overcome these shortcomings, we have developed a new type of low-cost fiber diffusers based on a graded-scattering method in which the optical scattering of the fiber diffuser is gradually increased as the light travels. The graded scattering can compensate for the optical attenuation and provide relatively uniform light emission along the diffuser. We performed Monte Carlo numerical simulations to optimize several key design parameters, including the number of scattering segments, scattering anisotropy factor, divergence angle of the optical fiber, and reflective index of the surrounding medium. These optimized parameters collectively result in uniform light emission along the fiber diffuser and can be flexibly adjusted to accommodate different applications. We fabricated and characterized the prototype fiber diffuser made of agarose gel and intralipid. Equipped with the new fiber diffuser, we performed thorough proof-of-concept studies on *ex vivo* tissue phantoms and an *in vivo* swine model to demonstrate the deep-imaging capability (~10 cm achieved *ex vivo*) of photoacoustic tomography. We believe that the internal light delivery via the optimized fiber diffuser is an effective strategy to image deep targets (*e.g.*, kidney) in large animals or humans.

## I. Introduction

Photoacoustic tomography (PAT), a hybrid imaging modality that combines optical excitation with ultrasound detection, can provide relatively deep penetration in biological tissues with acoustically determined spatial resolutions. In PAT, optical energy (scattered or not) absorbed by tissues is partially converted via the photoacoustic effect into pressure waves, which are subsequently received by an ultrasonic transducer or transducer array to form a tomographic image of the initial optical energy deposition [1]. PAT has been applied in a variety of biomedical applications to provide structural, functional, and molecular information [2-7]. In particular, taking advantage of a high sensitivity to blood (flowing or static), PAT has played an increasingly important role in studying vascular structure, oxygen saturation of hemoglobin [6, 8], blood flow [9, 10], and metabolic rate of oxygen [4, 11] as well as in interventional procedures [12-14]. We recently demonstrated real-time monitoring of shockwave-induced vascular injury during shockwave lithotripsy (SWL), a treatment that uses high-energy shockwave pulses to break stones in the kidney or peripheral organs [15]. We have demonstrated that PAT can potentially provide urologists with feedback of the vascular injury, which cannot yet be achieved with clinical imaging modalities (*e.g.*, fluoroscopy and B-mode ultrasound imaging) [16]. In our previous experiments using PAT, we adapted the conventional illumination design with a bifurcate optical guide flanking the linear ultrasonic transducer array [17, 18]. While this external illumination strategy works for small animal models, it does not work for studies on human kidneys that are more than 7 cm deep due to the strong optical attenuation of the tissue [19]. A limited imaging depth is the major hurdle for translating PAT to clinical SWL applications.

Although kidneys are deeply seated underneath the skin surface, they are connected directly to the urinary tract, which

This work was supported in part by National Institute of Health (R01EB028143, R01 NS111039, R01 NS115581, R01GM134036, R21 EB027304, R43 CA243822, R43 CA239830, R44 HL138185); Duke MEDx Basic Science Grant; Duke Center for Genomic and Computational Biology Faculty Research Grant; Duke Institute of Brain Science Incubator Award; American Heart Association Collaborative Sciences Award (18CSA34080277). (*Corresponding author: Junjie Yao*).

M. Li, T. Vu, and J. Yao are with the Department of Biomedical Engineering, Duke University, Durham, NC 27708, USA. (e-mail: junjie.yao@duke.edu).
G. Sankii and P. Zhong are with the Department of Mechanical Engineering and Materials Science, Duke University, Durham, NC 27708, USA.
B. Winship, K. Boydston, and R. Terry are with the Department of Surgery, Duke University School of Medicine, Durham, NC 27708, USA.

can be accessed from outside the body. An internal-illumination strategy is suitable to deliver light to the inside of the kidneys through the ureter. Photons can access the renal vasculature without experiencing otherwise strong attenuation by intervening tissues between the skin and kidney. Several internal-illumination PAT studies have been performed for brain imaging [20, 21], prostate imaging [22-24], nerve detection [13, 25], placenta imaging [26], and catheter and needle tracking in interventional procedures [27, 28]. In the reported internal-illumination designs, light emits either directly from the conical or flat fiber tips (*i.e.,* forward-viewing illumination) or from one side of the fiber by using an angled fiber tip or a 45° mirror (*i.e.,* side-viewing illumination). In SWL, renal injury often manifests as blood vessel hemorrhage within the SW focus, which has a dimension of a few centimeters. To cover such a large tissue volume, a side-viewing illumination is preferred to the forward-viewing illumination. However, the reported internal-illumination designs may not be able to provide uniform illumination over a large tissue volume within the shockwave focus. A novel light delivery strategy that can release light uniformly is still needed for our application.

We have previously demonstrated the concept of internal-illumination PAT using a customized radial-emission fiber diffuser, which enabled illumination of a relatively large tissue volume and achieved an imaging depth of ~7 cm in tissue phantoms [29]. Radial-emission fiber diffusers are commonly used in biomedical applications such as photothermal and photodynamic therapies. To make a radial-emission fiber diffuser, the jacket and cladding of an optical fiber are removed, and the surface of the fiber core is roughened to scatter the light out [30, 31]. The fiber core can also be coated with a layer of optically scattering particles to further homogenize the light emission [32, 33]. However, the radial-emission fiber diffuser has several drawbacks: (1) the light emission along the diffuser is largely determined by the roughness of fiber surface, which is technically challenging to precisely control in fabrication; (2) the light emission decreases exponentially along the diffuser and thus the resultant illumination is not uniform; and (3) without the protection of cladding and jacket, the diffuser is exposed directly to the surrounding medium and thus can be easily contaminated and damaged. Efforts have been made to overcome the aforementioned issues. For example, a focused laser beam can be applied to imprint gratings or drill micro-holes on the fiber core surface to precisely control the roughness [34-36]; the incident light angle can be carefully controlled to achieve more uniform light emission along the fiber [31]; the exposed glass fiber core can be etched with hydrofluoric acid to completely remove the cladding [37, 38]; and the diffuser tip can be enclosed by an additional glass envelope to protect the bare fiber core [39]. Nevertheless, these measures require a complex fabrication process, which significantly increases the cost of and/or degrades the performance (*e.g.*, reduced light coupling efficiency) of the fiber diffuser and thus are not ideal for internal-illumination PAT.

In this work, we have developed a new fiber diffuser based on the graded-scattering method. The light emission of the new fiber diffuser can be precisely tuned by adjusting the optical properties of the scattering materials in a simple and cost-effective way. We derived the mathematical model of the graded scattering and investigated key factors that affect the diffuser's performance through numerical simulations. We performed tissue phantom experiments using the optimized diffuser in which *ex vivo* pig kidneys were imaged to mimic the large organ depth. Finally, the validated imaging system was applied to an *in vivo* swine model. These studies collectively demonstrate the deep imaging capability of the internal-illumination PAT system enabled by the new graded-scattering-based fiber diffuser.

## II. METHOD

### A. Principle of the graded-scattering based fiber diffuser

If the fiber diffuser has a constant scattering coefficient such as in the commercial radial-emission diffuser we utilized previously [21], the light emission decreases exponentially along the diffuser, as described by Beer's Law. Thus, a large percentage of optical energy is emitted at the proximal end (defined as the end facing the output tip of the optical fiber) of the diffuser. To address this issue, we adapted a graded-scattering strategy that can compensate for the light attenuation along the diffuser and achieved more uniform light emission [35]. To do so, we used weakly-scattering medium at the proximal end of the diffuser that allows more photons to travel forward and gradually increase the scattering along the diffuser to strengthen the light emission. Ideally, the scattering coefficients should change continuously and smoothly along the diffuser, which, however, is a technologically challenging process. To simplify the fabrication, we assembled a series of cylindrical-shaped medium segments with increasing scattering coefficients (**Fig. 1**(a)). The segments have equal lengths, and the scattering coefficient within each segment is uniform. When the segments are short and the scattering coefficients are precisely controlled, the light emission along the diffuser can be approximately uniform as modeled below.

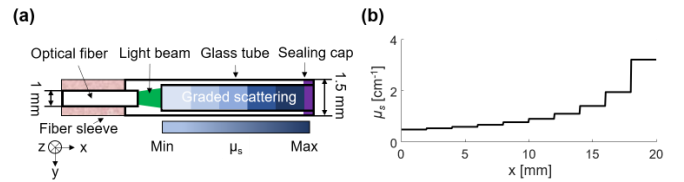

**Fig. 1** (a) Principle of the graded-scattering based fiber diffuser made of segmented scattering medium. (b) Computed scattering coefficients along the diffuser with a total length of 2 cm, a total number of 10 segments, and a residual transmission rate of 10%.

We define the total optical energy coupled into the fiber diffuser as $E_0$. The diffuser has a total length of L with N segments. T is the percentage of light that can eventually survive the diffusion and exit from the diffuser's distal end. The radial light emission is assumed to be cylindrically symmetric along the diffuser. We also assume that the diffuser has negligible optical absorption at the near-infrared wavelengths. For each segment, the total emitted optical energy due to scattering can be expressed as

$$\Delta E = (1-T)\frac{E_0}{N} \quad (1).$$

Thus, the optical energy entering the *n-th* segment is $E(n) = E_0 - (n-1)\Delta E$. Assuming the scattering coefficient of the *n-th* segment is $\mu_s(n)$, we get the scattered optical energy by the *n-th* segment as

$$\Delta E = E(n)\left(1 - exp\left[-\mu_s(n)\frac{L}{N}\right]\right) \quad (2).$$

We assume that all scattered photons eventually exit the *n-th* segment, which, as we will discuss below, can be adjusted by optimizing the scattering anisotropy factor of the scattering medium and the relative refractive index of the surrounding medium.

Substituting (1) into (2), we get:
$$\mu_s(n) = -\frac{N}{L}ln(1 - \frac{\Delta E}{E(n)}) = -\frac{N}{L}ln(1 - \frac{1-T}{N-(n-1)(1-T)}) \quad (3).$$

Based on (3), we can calculate the scattering coefficient of each segment once we have determined the key parameters, including the total diffuser length *L*, the number of segments *N*, and the residual transmission rate *T*.

### B. Optimization of the diffuser parameters

To optimize the diffuser's performance, we used Monte Carlo (MC) simulations (Tracepro, Lambda, MA, USA) to investigate the impacts of four key parameters, including the number of segments, the scattering anisotropy of the scattering medium, the relative refractive index of the surrounding medium, and the beam divergence of the optical fiber.

*1) Diffuser geometry*

The geometry of the diffuser, including the diameter and total length, must be carefully designed according to the application. For example, to illuminate a pig kidney, the fiber diffuser has to fit into the channel of an ureteroscope, which limits the total diameter of the diffuser to < 1.5 mm. Additionally, to couple high optical energy into the diffuser, a large fiber core is highly desired. Therefore, we chose a silica optical fiber with a core diameter of 1 mm. After the protective sleeve was removed, the total diameter of the core and the cladding is ~1 mm. The fiber tip was then inserted into a glass tube with an inner diameter of ~1.1 mm and an outer diameter of 1.3 mm. The same glass tube also houses the fiber diffuser.

We chose a total length of 20 mm for the fiber diffuser because of three major considerations: (1) the detection aperture size of the ultrasonic transducer array is ~40 mm in our internal-illumination PAT, so the length of the diffuser should be < 40 mm; (2) the 20 dB optical diffusion length at 1064 nm in soft tissue is ~10 mm, so the diffuser length should be no less than 20 mm to provide a sufficient illumination volume; and (3) since the optical fiber may have to be inserted through the endoscope channel and the fiber diffuser is rigid, the length of the diffuser has to be as short as possible to maintain the flexibility.

A low residual transmission rate T is preferred to maximize the total light emission. We had empirically set T to 10% in our diffuser design, and further reductions do not have a large impact on the diffuser's performance. By computing the scattering coefficient of each segment according to Equation (3), we simulated the diffuser performance with different numbers of segments (N) but with the same total length (L) of 20 mm. Increasing the number of segments reduces the difference in the scattering coefficients between adjacent segments and thus improves the uniformity of the light emission. However, it is challenging to fabricate the short diffuser segments. In our study, 5, 10, 15, and 20 segments were simulated, and their corresponding illumination patterns were compared. From the MC simulation results, the normalized light emission intensity was computed along the fiber diffuser with different segment numbers. To quantify the uniformity of the light emission, we also calculated the mean, standard deviation (STD), and coefficient of variation (CV) of the normalized emission intensity profiles.

*2) Scattering anisotropy factor*

The impact of the scattering anisotropy on the diffuser's performance is usually neglected [37, 40]. Ideally, with an anisotropy factor near zero, the photons are scattered uniformly in all directions, and most of the scattered photons can contribute to the diffuser's local emission. However, when the scattering medium's anisotropy factor is not negligible, some scattered photons still travel forward and do not contribute to the local diffuser emission. For example, the anisotropy factor of the agar-intralipid phantom is ~0.47 [41] and varies slightly with the intralipid concentration. We simulated eight anisotropy factors: 0, 0.15, 0.30, 0.45, 0.60, 0.70, 0.80, and 0.90 and computed the STD and CV of the normalized light emission profiles.

*3) Refractive index of the surrounding medium*

The light emission rate of the fiber diffuser also depends on the refractive index of the surrounding medium outside of the glass tube that houses the diffuser. The glass tube has a refractive index (*n*) of 1.45. The refractive index of the surrounding medium determines the critical angle of total internal reflection, which has an impact on the scattered photons that can exit the diffuser locally. To study the impact of the surrounding medium's refractive index, we simulated the diffuser's performance in water (*n*=1.33) and in air (*n*=1.0).

*4) Beam divergence angle of the optical fiber*

The emission rate of the diffuser also depends on the beam divergence angle of the attached optical fiber. A small divergence angle is preferred because it helps trap photons inside the diffuser, particularly at the proximal end. Our off-the-shelf optical fiber has a 15˚ divergence angle, which may lead to a higher emission rate at the proximal end. Therefore, we simulated the impact of the divergence angle on the diffuser's performance in air and in water.

*5) Monte Carlo Simulation*

We used Tracepro (Lambda, MA, USA) for the MC simulations that mimicked our practical diffuser fabrication. In the MC simulations, identical cylinder segments with 1 mm diameter were stacked to build the numerical diffuser. The total length of the diffuser was 20 mm and the length of each cylinder was determined by the number of segments. Each segment had the same mass density and refractive index as water. Two concentric cylindrical shells with a wall thickness of 0.15 mm were aligned co-axially with the diffuser. The inner cylindrical shell (inner diameter: 1 mm) simulated the glass capillary that housed the diffuser, and the outer cylindrical shell (inner diameter: 1.3 mm) simulated the working environment (water or air). The light emission intensity was measured along the surface of the outer cylindrical shell. A diverging light beam, simulating the output of a multimode optical fiber, was launched towards the proximal end of the diffuser from 1 mm distance. A total of 76,028 incident light rays were simulated with a total power of 1 Watt.

*C. Fabrication and validation of the diffuser*

Based on the MC simulation results, we fabricated prototype diffusers using agar mixed with intralipid. A CCD camera was used to capture the light emission profile of the diffusers.

*1) Fabrication of the graded-scattering fiber diffuser*

We followed four major steps to fabricate the fiber diffuser. (1) Based on the theoretical model and MC simulation results, we determined the number of diffuser segments and the scattering coefficient of each segment; (2) We made each diffuser segment using a mixture of agar (Cat# A7002, Sigma-Aldrich, MO, USA) and intralipid (Cat# I141, Sigma-Aldrich, MO, USA). One gram of agar powder mixed with 100 mL of deionized water was heated by a microwave oven for ~1 minute until the agar powder was fully dissolved. Stock intralipid solution was added to the agar solution to adjust the scattering coefficient and the anisotropy factor of the final mixture. To achieve the desired scattering coefficient of each diffuser segment, the intralipid concertation was determined based on previously described study [41]. The final scattering coefficients and corresponding intralipid concentrations are shown in **TABLE I**. (3) A glass tube with an inner diameter of ~1 mm was used to house the agar-intralipid mixture. We used a syringe with a 22-gauge needle to inject the agar-intralipid mixtures into the glass tube, following the segment sequence determined in Step (1). For each segment, 1.57 μL mixture was injected, which quickly solidified inside the glass tube. This procedure was repeated until all the segments were injected and completely solidified. The empty portion of the glass tube was removed by using a circle glass cutter. The distal end of the diffuser was then sealed with silicone glue. (4) Finally, we removed the sleeve of the output tip of a multi-mode optical fiber (Flexiva, Boston Scientific, MA, USA) and fixed the fiber tip at 1 mm distance to the proximal end of the diffuser using UV glue (NOA 61, Thorlabs, NJ, USA). We stored the fiber diffuser at 4 °C to improve the shelf life.

**TABLE I.** PARAMETERS OF DIFFUSER SEGMENTS

| Seg # [a] | $\mu_s$ (cm$^{-1}$) | g | Intralipid (μL/100 mL) [b] |
|---|---|---|---|
| S1 | 0.220 | 0.478 | 18.50 |
| S2 | 0.260 | 0.478 | 21.87 |
| S3 | 0.300 | 0.478 | 25.23 |
| S4 | 0.383 | 0.477 | 32.22 |
| S5 | 0.460 | 0.477 | 38.71 |
| S6 | 0.650 | 0.477 | 54.73 |
| S7 | 0.850 | 0.477 | 71.61 |
| S8 | 1.500 | 0.477 | 126.63 |
| S9 | 2.100 | 0.476 | 177.62 |
| S10 | 4.981 | 0.474 | 425.18 |

[a] The segment number from the proximal end of the diffuser.
[b] The stock intralipid volume per 100 mL agar solution.

*2) Ex vivo and in vivo experimental setup*

Our internal-illumination PAT system includes a 128-channel programmable ultrasound scanner (Vantage 128, Verasonics, USA), a linear ultrasonic transducer probe (L7-4, Philips, USA; Central frequency: 5.2 MHz), and a pulsed Nd:YAG laser (Q-smart 850, Quantel Laser, USA; Wavelength, 1064 nm). The laser beam was coupled into the multi-mode fiber via a convex lens, with a pulse energy of ~15 mJ. The ultrasound scanner was synchronized with the laser firing at 10 Hz. A bandpass filter (cutoff frequencies: 4 MHz and 7 MHz) was applied to the raw channel data before the image reconstruction.

We quantified the spatial resolutions of the internal-illumination PAT system by imaging a black hair (diameter ~100 μm) immersed in water as a line target. The distance between the hair and the ultrasound probe surface was ~80 mm. The full-width-at-half-maximum (FWHM) of the reconstructed hair profile was measured as the spatial resolution. We also measured the maximum penetration depth of the fiber diffuser by imaging a 2-mm (diameter) tube filled with whole bovine blood (Quad Five, MT, USA). The diffuser was placed at different depths from the tube, which was immersed in an optically scattering medium made of gelatin and 1% intralipid (v/v). The estimated reduced scattering coefficient of the medium was ~7 cm$^{-1}$ at 1064 nm [41]. A total of 300 PA images were averaged at each depth, and the signal-to-noise ratios (SNRs) were quantified. Signal averaging was performed to improve the SNR, which is commonly used for imaging deep tissues. For our targeted clinical application of shockwave lithotripsy, signal averaging is acceptable, since the shockwave repetition rate is only 1 Hz and the vascular damage is a fairly slow process. However, to perform signal averaging for *in vivo* applications, motion artifacts need to be minimized, for example, by respiratory-gated data acquisition [42].

We validated the performance of the fiber diffuser in both *ex vivo* and *in vivo* studies. For the *ex vivo* study, we prepared a pig kidney from a local farm, with a thickness of ~2 cm and a reduced scattering coefficient of ~11.2 cm$^{-1}$ at 1064 nm [43, 44]. Because the kidney's natural blood vessels were not perfused after extraction, a plastic tube filled with bovine blood was wrapped in a spiral pattern and placed on top of the kidney to mimic the blood vessels. The kidney and blood-filled tube were then covered by a 6-cm-thick layer of fresh chicken breast tissue. The fiber diffuser was inserted into the kidney through the remaining urinary tract, approximately 1 cm from the blood tube. The ultrasonic transducer probe was placed on the top of the chicken breast tissue to detect the PA signals.

As the PA signals are generated from targets deeply seated inside the tissue and detected on the tissue surface, the ultrasound attenuation can be significant. In particular, muscle and skin are both strong acoustic attenuators [45]. We further investigated the impact of ultrasound attenuation by placing chicken breast tissue and pig skin on top of the pig kidney. To mimic the acoustic attenuation by skin and muscle, PA images were acquired through a layer of pig belly skin tissue (35 mm thick) and chicken breast tissue of various thickness (0 mm, 15 mm, 35 mm, 50 mm, and 65 mm), which were placed on top of a pig kidney with a blood tube in between.

As a proof-of-concept *in vivo* study, we used a female piglet with a weight of ~50 kg because the size and anatomy of the pig kidney are similar to that of an adult human [46]. All experiments were approved by the Institutional Animal Care and Use Committee of Duke University. Following anesthetization, fluoroscopy was performed to guide the ureteroscope to the lower pole of the kidney. The fiber diffuser was then advanced through the ureteroscope's catheter channel to the lower pole, as confirmed by the fluoroscopy. The pig kidney was then surgically exposed, and the ultrasound probe was placed directly on top of the kidney. Both B-mode ultrasound and PA images of the pig kidney were acquired.

*3) PAT image reconstruction*

A delay-and-sum (DAS) method was applied to reconstruct all of the PAT images [47, 48]. The size of the reconstructed PA image was 40 mm (lateral) by 110 mm (axial), with a pixel size of 65 μm.

## III. EXPERIMENTAL RESULTS

### A. Optimization of the fiber diffuser

**Figure 2** shows the simulated impact of the total number of diffuser segments on the light emission. The scattering coefficients of the diffuser with five segments were calculated, and the emission distribution was simulated (**Fig. 2**(a)). In this case, each segment was 4 mm long, and the scattering coefficients increased sharply. The light emission fluctuated greatly along the diffuser. For the diffuser with 10 segments, the overall emission profile was more uniform along the diffuser, especially at the distal end (**Fig. 2**(b)). No substantial differences in the emission profiles were observed among 10-, 15- and 20-segment diffusers, all of which showed relatively uniform emission (**Figs. 2**(c-d)). Our quantitative results, including the mean, stand deviation (STD) and coefficient of variation (CV) of normalized optical intensity distribution, are shown in **TABLE II**. The 5-segments diffuser has the largest STD and CV, while the performances of the 10-, 15-, and 20-segments diffusers were similar. Therefore, we chose to implement the 10-segments design, which offered the best compromise between emission uniformity and fabrication complexity.

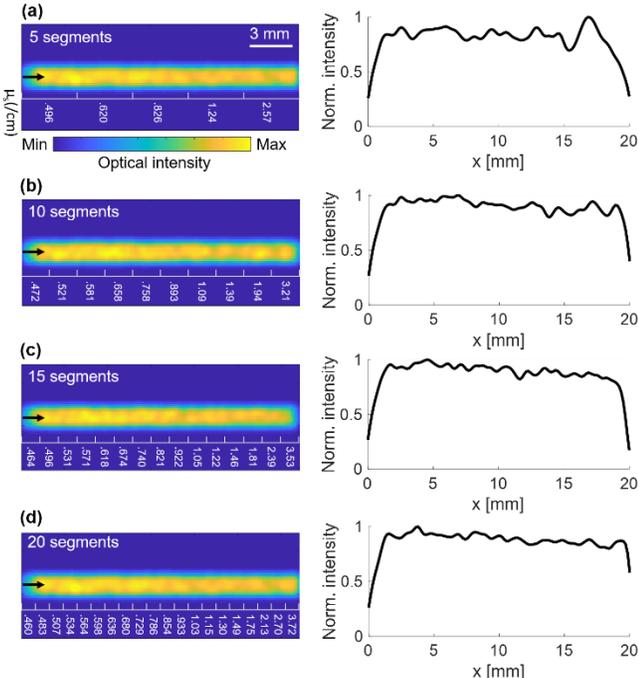

**Fig. 2** Simulated light emission distributions and profiles along the diffuser with (a) 5 segments, (b) 10 segments, (c) 15 segments, and (d) 20 segments. The corresponding scattering coefficients (cm$^{-1}$) are marked below each emission distribution.

**TABLE II.** NORMALIZED OPTICAL INTENSITY ALONG THE DIFFUSER WITH DIFFERENT NUMBER OF SEGMENTS

| N [a] | Mean | STD | CV (%) |
|---|---|---|---|
| 5 | 0.836 | 0.071 | 8.420 |
| 10 | 0.911 | 0.049 | 5.430 |
| 15 | 0.903 | 0.049 | 5.460 |
| 20 | 0.889 | 0.042 | 4.730 |

[a] Number of segments in the MC simulation.

The impact of the anisotropy factor was also simulated. The light emissions of the diffuser with four different anisotropy values are shown in **Fig. 3**. When the anisotropy factor is 0, 0.3, or 0.6, the light emission is nearly uniform along the diffuser (**Figs. 3**(a-c)). When the anisotropy factor is increased to 0.9 (**Fig. 3**(d)), less light is emitted at the proximal end, as most scattered photons still travel forward along the diffuser, and thus Eq. (2) is no longer valid. The STDs and CVs of the intensity profiles with different anisotropy factors are shown in **Figs. 3**(e-f). From the simulation results, we conclude that an anisotropy factor between 0.4 and 0.7 (STD ~0.05, CV ~5%) is most suitable for constructing our fiber diffuser.

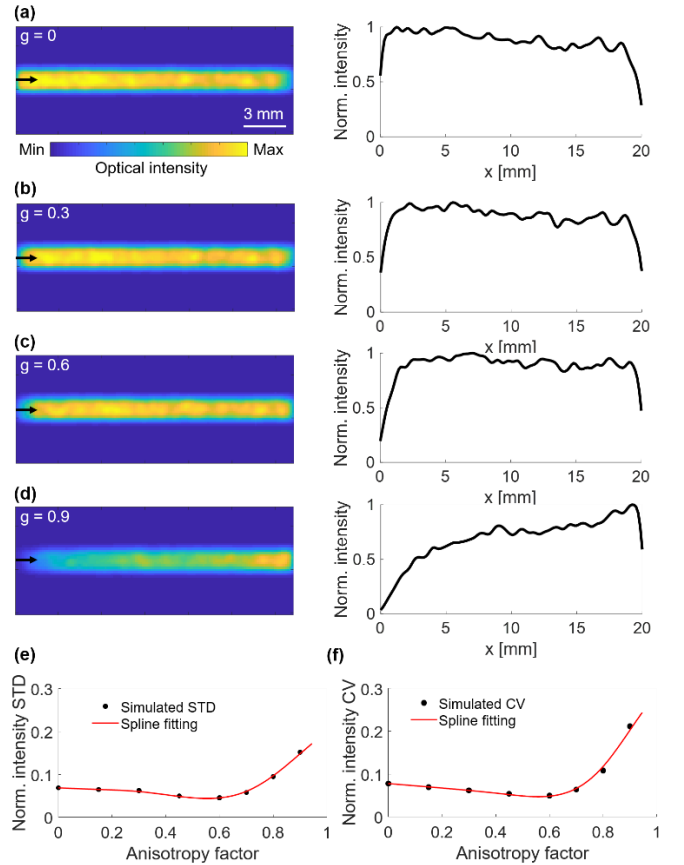

**Fig. 3** Simulated light emission distributions and profiles along the diffuser with scattering anisotropy factors of (a) 0, (b) 0.3, (c) 0.6, and (d) 0.9. (e) The STD of normalized optical intensity along the fiber diffuser with different anisotropy factors. (f) The CV of normalized optical intensity along the fiber diffuser with different anisotropy factors. Black dots represent the simulated values and red lines represent the spline fitting.

**Figure 4** shows the impact of the refractive index of the surrounding medium on the performance of the diffuser. All of the intensity profiles were normalized by the maximum intensity in **Fig. 4**(d), and the quantitative comparisons are shown in **TABLE III**. With the collimated input beam, the fiber diffuser in air has low emission rate along the diffuser, and the distribution is highly inhomogeneous (**Fig. 4**(a)). The light emission rate at the proximal end is 39% lower than the rate at the distal end. On the contrary, when the fiber diffuser is in water, the overall light emission is higher and more uniform (**Fig. 4**(b)). The low and non-uniform emission in air is due to the large refractive index mismatch between the glass tube and air, which leads to a critical angle of total internal reflection at the glass-air interface of ~43.6°. When the scattered light arrives at the glass-air interface with an approximately uniform angle distribution (*i.e.*, with a small scattering anisotropy factor), more than half of the photons are reflected back into the diffuser instead of exiting the local diffuser segment. In contrast, at the glass-water interface, with a critical angle of ~66.5°, the impact of total internal reflection is reduced, and thus more photons can exit the local diffuser segment.

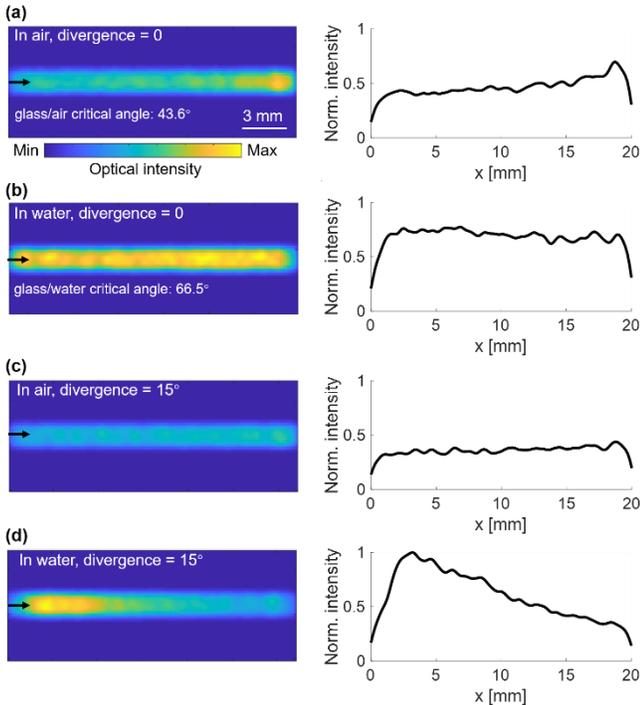

**Fig. 4** Simulated light emission distributions and profiles along the diffuser in air (a and c) and in water (b and d) with an input light beam divergence angle of 0° (a-b) or 15° (c-d).

In our expected applications using internal-illumination PAT, the fiber diffuser will be advanced into the kidney via the ureter filled with urine. Therefore, water is a more applicable surrounding medium than air for our current design. In other scenarios, when the diffuser is surrounded by air, such as in the esophagus or lungs, the scattering coefficients of segments must be adjusted to compensate for the stronger total internal reflection.

The impact of a 15° divergence angle on the light emission of the diffuser is shown in **Figs. 4**(c-d). The beam divergence has a large impact both in air and in water. When the diffuser is in air, the diverged beam improves the emission uniformity compared with the collimated beam, but the overall emission is still low. The total internal reflection at the glass-air interface traps more than 50% photons inside the diffuser. When the diffuser is in water, the diverged beam results in strong light emission at the proximal end, which quickly decreases along the diffuser. The light emission at the distal end is only ~30% of that at the proximal end. The larger the divergence angle, the quicker the light emission decreases along the diffuser.

**TABLE III.** SIMULATIONS OF NORMALIZED OPTICAL INTENSITY ALONG THE DIFFUSER WITH DIFFERENT SURROUNDING MEDIUM & INPUT BEAM DIVERGENCE

| Env. / Div. [a] | Mean | STD | CV (%) |
|---|---|---|---|
| Air/0° | 0.473 | 0.070 | 14.67 |
| Water/0° | 0.911 | 0.049 | 5.340 |
| Air/15° | 0.364 | 0.026 | 7.260 |
| Water/15° | 0.631 | 0.211 | 33.48 |

[a] Env.: Surrounding medium; Div.: Input beam divergence angle.

The impact of the beam divergence on the diffuser's light emission in water can be minimized by correcting the graded scattering coefficients based on the emission slope obtained from **Fig. 4**(d). The simulation results with the corrected scattering coefficients are shown in **Fig. 5**(a); of note, there is more uniform light emission along the diffuser. The fabricated prototype diffusers are shown in **Fig. 5**(b), with the intralipid concentration labelled for each segment. The light emission patterns of the prototype diffusers are shown in **Figs. 5**(c-d). The light emission profiles of the prototype diffusers show good agreement with the simulations.

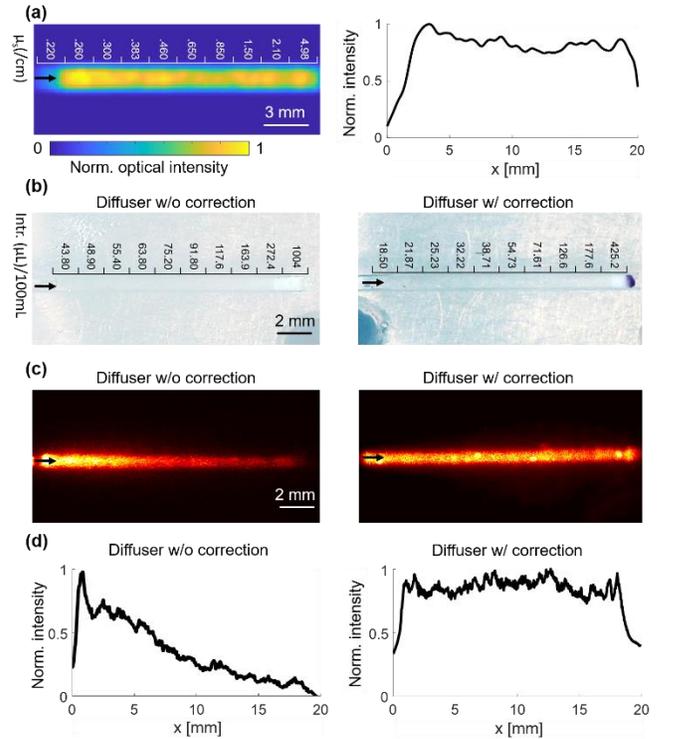

**Fig. 5** Comparison of light emission with and without correcting the input beam divergence. (a) Simulated light emission distribution and profile with beam divergence correction. The corresponding scattering coefficients are in the units of cm$^{-1}$. (b) Photos of prototype diffusers with and without beam divergence correction. The intralipid volume per 100 mL agar solution was marked above each segment. Intr.: Intralipid. (c-d) CCD camera images of light emission distribution (c) and profile (d) of the prototype diffusers with and without beam divergence correction.

*B. Internal-illumination PAT with the fiber diffuser*

The schematic of our internal-illumination PAT system is shown in **Fig. 6**(a). To characterize the system, the blood-filled tube was clearly imaged 3 cm away from the diffuser, demonstrating the large penetration of the diffuser. The SNR of the PA signals decreased with depth with a decay constant of ~10 dB/cm (**Figs. 6**(b-c)). The lateral and axial resolutions at 8 cm were 916 μm and 462 μm, respectively (**Fig. 6**(d)), consistent with our previous work [29]. The lateral and axial resolutions of the PAT system degrade with the imaging depth due to the decreased numerical aperture of the ultrasound transducer array and the signal's bandwidth, respectively.

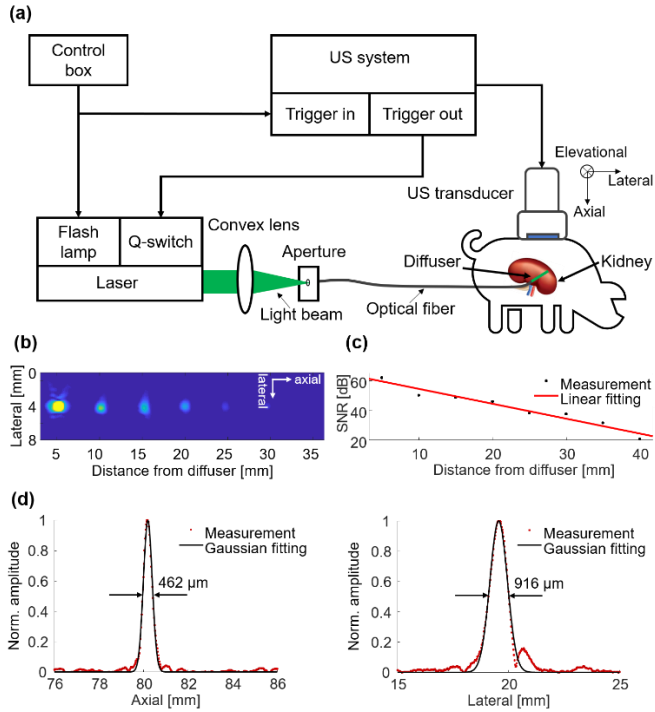

**Fig. 6** Characterization of the internal-illumination PAT system. (a) Schematics of the internal-illumination PAT system with the graded-scattering fiber diffuser. (b) Composite PAT image of a plastic tube filled with whole bovine blood embedded in an optically scattering medium, acquired at different depths from the fiber diffuser. (c) Measured (black dotted line) SNR of the tube as a function of depth. A linear fitting (red solid line) quantified the signal attenuation coefficient. (d) Spatial resolutions of the PAT system at a depth of ~80 mm from the ultrasonic transducer surface.

### C. Ex vivo and in vivo experiments

The *ex vivo* study, shown in **Fig. 7**(a), further validates the diffuser's performance in biological tissues. We conducted concurrent B-mode ultrasound and PA imaging of the tissue phantom, as shown in **Fig. 7**(b). Strong PA signals from the blood tube were detected. To demonstrate the superior imaging depth of the internal-illumination PAT, additional chicken breast tissue was stacked to achieve a total overlaying thickness of ~8 cm, as shown in **Fig. 7**(c). While the ultrasound image could not identify the tube, the PA image still has relatively high contrast from the blood tube. To image the structure of the blood tube (**Fig. 7**(d)), we scanned the ultrasound probe across the phantom. The top-view depth-projected and volumetric renderings of the PA results overlaid with the ultrasound image are shown in **Figs. 7**(e-f), respectively. In this experiment, reperfusion of the *ex vivo* kidney is preferred, but it is technically challenging. After extraction, the blood vessels inside the kidney were mostly collapsed or blocked. Reperfusion is more doable on freshly extracted kidneys [49].

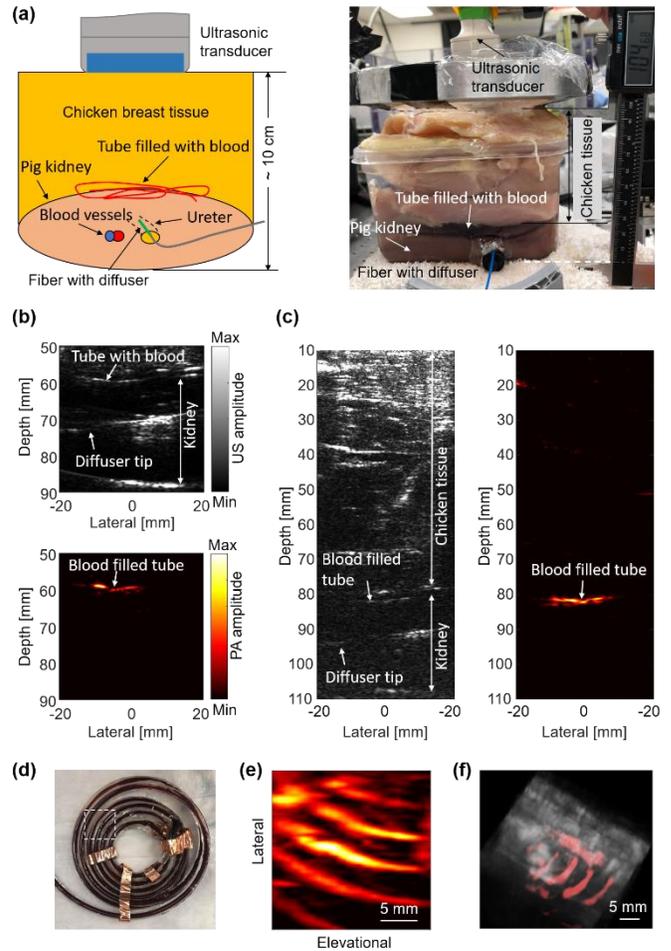

**Fig. 7** Validation of the graded-scattering fiber diffuser on *ex vivo* tissue phantoms. (a) Schematics and photograph of the *ex vivo* experimental setup. (b-c) B-mode ultrasound image and PA image of the blood-filled tube overlaid by a layer of fresh chicken breast tissue with a total thickness of (b) 6 cm and (c) 8 cm. (d) Photograph of the blood-filled tube. The white-dotted box indicates the imaged region. (e) Depth-projected PA image of the blood-filled tube. (f) Volumetric rendering of the PA image (shown in color) overlaid with the ultrasound image (shown in gray).

As shown in **Fig. 8**(a), the SNR of the PA signal from the blood tube was 37.9 dB when the tube was underneath 60-mm-thick chicken tissue only. The SNR decreased to 27.4 dB when the top layer chicken tissue was replaced with 35-mm-thick pig belly skin tissue (**Fig. 8**(b)). The PA image can clearly identify the blood tube even through 35 mm of pig belly skin tissue and 65 mm of chicken tissue, with an SNR of 17 dB. The SNR decay of the PA signal with increasing tissue thickness is shown in **Fig. 8**(c). The SNR decay constant is 3.5 dB/cm.

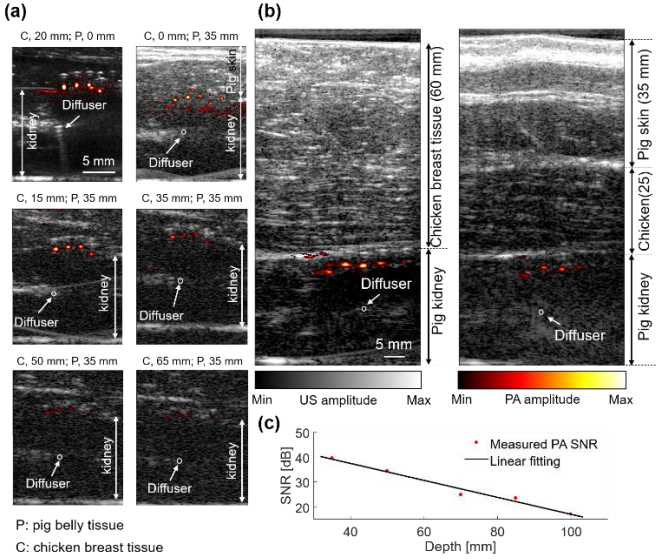

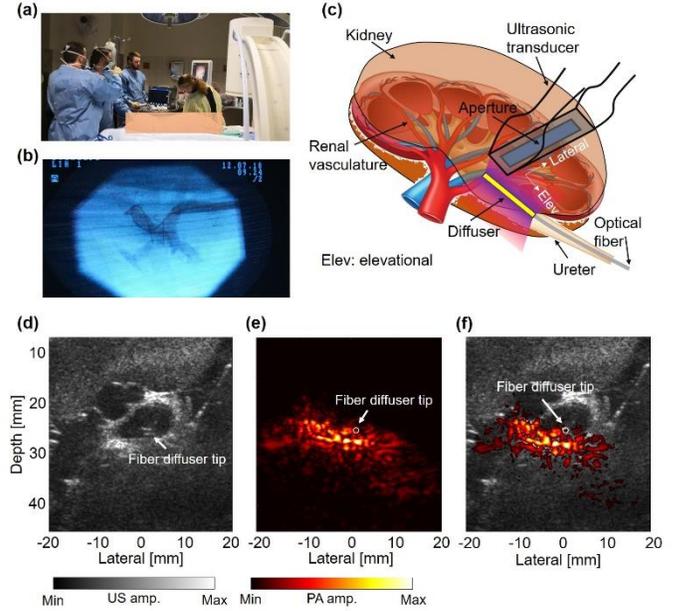

**Fig. 8** PA signal attenuation on *ex vivo* tissue phantoms. (a) PA images of the blood tube (shown in color) overlaid with B-mode ultrasound images (shown in gray), with different thickness of pig belly tissue (P) and chicken tissue (C). (b) Overlaid PA and ultrasound images with 60 mm chicken tissue only (left), or 35 mm pig belly skin tissue on top of 25 mm chicken tissue (right). (c) Measured (red dotted line) SNR of the PA signals of the blood tube as a function of the thicknesses of overlaying tissue. A linear fitting (black solid line) quantified the SNR decay coefficient.

**Fig. 9** Validation of the diffuser's performance on a swine model *in vivo*. (a) Photograph of the *in vivo* pig experiment setup. (b) Fluoroscopy image of the ureteroscope's catheter channel in the pig kidney. (c) Schematics of the imaging setup. (d) B-model ultrasound image of the lower pole of the exposed pig kidney *in vivo*. (e) Photoacoustic image of the lower pole of the exposed pig kidney *in vivo*. (f) Overlaid DAS-based PA image (shown in color) and B-mode ultrasound image (shown in gray) of the pig kidney.

**Figure 9**(a) shows the *in vivo* swine study setup. The fluoroscopy image of the lower renal pole is shown in **Fig. 9**(b), in which the optical fiber advanced from the ureteroscope can be clearly seen. The relative positions of the kidney, the ultrasonic transducer probe, and the fiber diffuser are illustrated in **Fig. 9**(c). **Figure 9**(d) shows a B-mode ultrasound image of the lower pole of the pig kidney, where the diffuser tip was also visible. **Figure 9**(e) shows the PA image of the same kidney section. The overlaid PA and ultrasound image of lower pole is shown in **Fig. 9**(f). The PA image using DAS shows majorly blood vessel signals around the diffuser, which, however, suffers from the limited-view and limited-bandwidth artifacts by using a linear ultrasound probe. Nevertheless, the *in vivo* results demonstrate the feasibility that laser light can be delivered by our newly designed fiber diffuser through the ureter to illuminate the kidney.

## IV. Discussions

The total delivered laser pulse energy via the fiber diffuser is one of the most important factors that affect the signal strength and penetration depth of the internal-illumination PAT. Higher SNRs can be achieved as more light is delivered. Since the diameter of the optical fiber is limited to 1 mm to go through the ureter, the maximum laser pulse energy that can be delivered is also limited. Given a laser pulse width of 10 ns, the theoretical damage threshold of the fused silicon fiber is only ~8 mJ at 1064 nm [50]. In practice, the optical energy delivered into the diffuser was ~15 mJ, resulting in an optical fluence of 23.8 mJ/cm$^2$ at the diffuse surface, which is much lower than the ANSI safety limit at 1064 nm on the skin (~100 mJ/cm$^2$). Because the fiber's damage threshold is proportional to the laser pulse width, one possible solution is to apply a longer pulse width and a larger pulse energy with a slight cost of spatial resolution, as shown by the simulation results in **Supplementary Fig. 1** [51]. For example, the resultant PA signal amplitude decreases 3.2-fold when the laser pulse width is increased from 10 ns to 150 ns. However, with a 15-times longer pulse width, the fiber's damage threshold is also increased by 15 times compared with the theoretical value, and a pulse energy of ~120 mJ can be delivered.

The laser safety limit inside the tissue should be more relaxed than that on the skin, because unlike skin, internal tissue does not have melanin pigment, which has much stronger absorption than hemoglobin. To apply internal-illumination PAT to the study of the kidney, the fiber diffuser's geometry must accommodate the dimensions of the kidney. Our *in vivo* swine experiment demonstrates that the current 20 mm diffuser with a 1-mm diameter is sufficiently flexible for kidney imaging. For other applications in different organs, the diffuser design should be optimized based on the organ's dimensions and accessibility.

Since an NIR wavelength (1064 nm) is used to achieve large penetration, potential heating of the water may alter the diffuser's optical properties. Given that the optical absorption coefficient of water at 1064 nm is ~0.145 cm$^{-1}$, within the thermal relaxation time of the diffuser (~1.7 seconds), the accumulated temperature rise inside the diffuser is ~0.58 °C, which is unlikely to change the diffuser's optical properties [52, 53].

We believe the *in vivo* PA signals at 1064 nm were from blood vessels inside the kidney instead of from the fiber diffuser. We specifically design our fiber diffuser so it has negligible absorption by all components: glass, intralipid, and agar. Our phantom experiments did not show signals from the fiber diffuser but only signals from the blood-filled tubes. The fiber itself should not generate PA signals since we removed a large portion of its sleeve, as detailed in the diffuser fabrication process. All the results are consistent with our previous publications showing that the major absorber at 1064 nm inside the biological tissues is hemoglobin [54, 55].

The quality of the *in vivo* PAT images is relatively poor, mainly because the linear ultrasound probe has limited view and limited detection bandwidth. Those artifacts may distort the reconstructed images. We believe that illuminating a large portion of the tissue is significantly more complex with denser structures, than that with a smaller illuminating area. The linear ultrasound probe is known to result in limited-view and limited-bandwidth artifacts for dense targets in photoacoustic imaging [56-59]. We will try different image reconstruction and data processing methods to improve the image quality. For example, the PA imaging quality can be improved by using a matrix transducer probe as well as model-based [60-63] or deep learning image reconstruction methods [64-66]. We have also explored our recently developed reconstruction approach based on the Wasserstein Generative Adversarial Network with gradient penalty (WGAN-GP) to improve the image quality (Supplementary **Fig. 2**) [67]. However, since we do not have ground truth for the *in vivo* study, the *in vivo* fidelity of the WGAN-GP method needs more validation. The current PAT setup does not have a co-current imaging modality that can image the blood vessels as a validation. Possible validations include contrast-enhanced ultrasound imaging or Doppler ultrasound imaging of the blood vessels.

## V. Conclusions

In this work, we have presented the design, fabrication, and proof-of-concept application of an optimized fiber diffuser for internal-illumination PAT. The fiber diffuser with graded scattering can achieve relatively uniform light emission and a large illumination volume. We first optimized the fiber diffuser parameters using Monte Carlo simulations and then fabricated a prototype fiber diffuser using a mixture of agar gelatin and intralipid; the diffuser has a total length of 20 mm, a total of 10 segments, an anisotropy factor of 0.47, and a fiber divergence angle of 15°. The performance of the fiber diffuser in air vs water was compared. The impact of input beam divergence was also corrected. The optimized fiber diffuser was applied in our internal-illumination PAT for *ex vivo* and *in vivo* studies. Our *ex vivo* results demonstrate a ~10-cm imaging depth in tissue phantom, and the *in vivo* pig experiment shows the feasibility of imaging the kidney vasculature with illumination from inside. In clinical practice, SWL of kidney stones often induces hemorrhage, which can be monitored by analyzing the changes in photoacoustic images, as demonstrated in our previous study [18]. Therefore, our next step is to apply the developed internal-illumination photoacoustic imaging on swine models to monitor renal hemorrhage during SWL. With the new fiber diffuser, we expect that internal-illumination PAT can find a wide range of deep-tissue applications in both fundamental research and clinical practice. For example, we can study the underlining correlation between the SWL-induced cavitation and the renal vascular injury on animal models.

## Acknowledgment

The authors thank Dr. Caroline Connor for editing the manuscript and Dr. Chengbo Liu for technical support on Tracepro.

## Supplementary figures

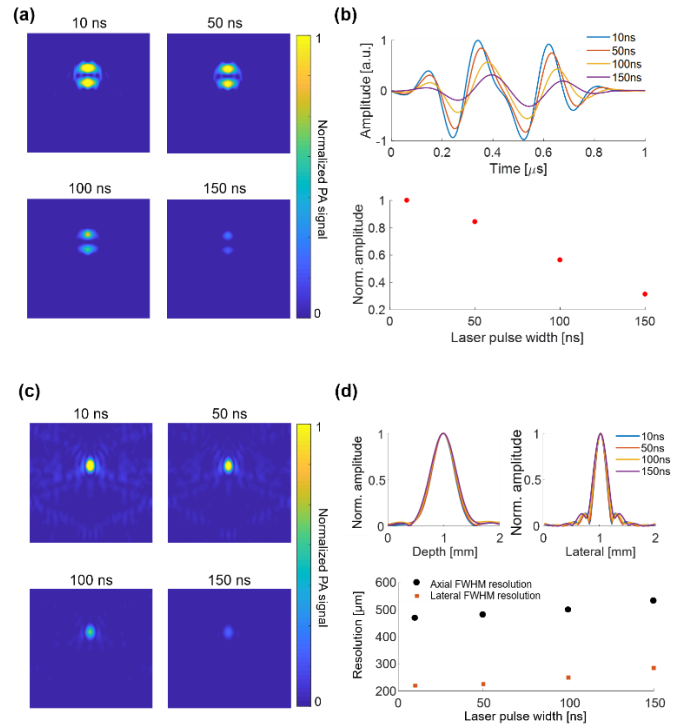

**Supplementary Fig. 1**. The PA signal and spatial resolution dependence on the excitation laser pulse width. (a) Simulated PA images of a 400-μm-diameter blood vessel, with an excitation laser pulse width of 10 ns, 50 ns, 100 ns, and 150 ns. The ultrasound detection has a central frequency 5 MHz and a 75% bandwidth. (b) The resultant PA signal profile and peak amplitude with the excitation laser pulse width ranging from 10 ns to 150 ns. (c) Simulated PA image of a point target (40-μm in diameter) with an excitation laser pulse width of 10 ns, 50 ns, 100 ns, and 150 ns. (d) The resultant spatial resolution measurements with the excitation laser pulse width ranging from 10 ns to 150 ns. A slight decrease in resolution was observed when the pulse width is stretched by 15 times.

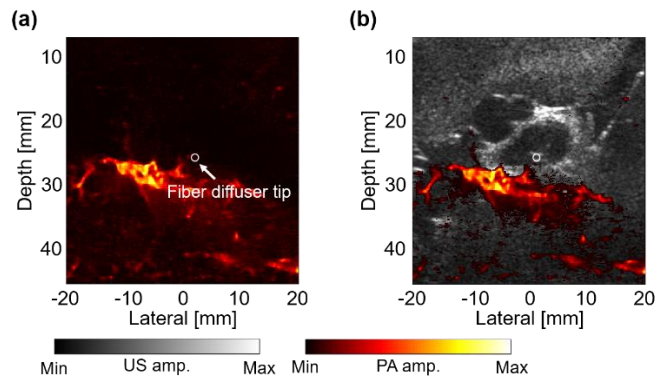

**Supplementary Fig. 2.** (a) PA image of the renal blood vessels using WGAN-GP. (b) Overlaid WGAN-GP-based PA image (shown in color) and B-mode ultrasound image (shown in gray) of the pig kidney.